\documentclass[aps,prx,twocolumn,nofootinbib,superscriptaddress]{revtex4-2}

\usepackage{amsmath,amssymb,amsthm}
\usepackage{braket}
\usepackage{xcolor,graphicx}
\usepackage[percent]{overpic}
\usepackage{bm,bbm}
\usepackage{booktabs}
\usepackage[colorlinks=true,linkcolor=blue!55!black,citecolor=blue!55!black,
            urlcolor=blue!55!black]{hyperref}
\usepackage{orcidlink} 

\newcommand\bfalpha{\boldsymbol{\alpha}}

\newcommand{\ii}{\mathrm{i}}
\newcommand{\ee}{\mathrm{e}}
\newcommand{\Tr}{\operatorname{Tr}}

\begin{document}
\title{Majorana Edge Modes Reshape Kibble--Zurek Scaling in Kitaev Honeycomb Ribbons}

\author{Hyesung Choi}
\affiliation{Department of Physics, Korea University, Seoul 02841, Republic of Korea}

\author{Antonio Mandarino}
\affiliation{International Centre for Theory of Quantum Technologies, University
of Gda\'nsk, prof. Marii Janion 4, Gda\'nsk, 80-309, Poland}

\author{Mahn-Soo Choi\,\orcidlink{0000-0001-6901-5233}}
\email{choims@korea.ac.kr}
\affiliation{Department of Physics, Korea University, Seoul 02841, Republic of Korea}

\begin{abstract}
Recent progress in programmable quantum simulators motivates exact
benchmarks for nonequilibrium dynamics in topological quantum matter.  We
study the Kitaev honeycomb model driven through its critical regions by a
linear ramp of the bond coupling $J_z$ over a ramp time $\tau$, comparing
zigzag and armchair ribbons with fully periodic tubes.  The production of
excitations exhibits a geometry-dependent finite-size crossover controlled
by $N_2/\sqrt{\tau}$, the ratio of the ribbon width to the Kibble--Zurek
length.  Periodic tubes obey the expected critical-line Kibble--Zurek
scaling and serve as the boundary-free benchmark.  Zigzag ribbons follow
the same scaling until defect production is exponentially suppressed by a
finite-width algebraic gap floor associated with the formation of Majorana
edge modes, whose occupation is pinned by inversion parity.  Armchair
ribbons instead support transient Majorana edge modes with exponentially
small gaps, producing width-parity-dependent defect plateaus: exactly one
defect per unit length for odd widths and an approximate, transient
plateau near one half for even widths.  The quench-generated entanglement
entropy and fermionic logarithmic negativity inherit the same
geometry-dependent structure, and the negativity further distinguishes
the coherence-borne entanglement of the Kibble--Zurek regime from the
population-borne entanglement of the edge-mode plateaus.  These results
identify boundary Majorana modes as the mechanism controlling departures
from ideal Kibble--Zurek dynamics and provide geometry-resolved benchmarks
for programmable simulations of topological matter.
\end{abstract}

\maketitle

\section{Introduction}

The dynamics of a many-body system driven across a continuous phase
transition provides a fundamental setting in which equilibrium criticality
meets intrinsically nonequilibrium physics.  Close to a critical point, the
characteristic relaxation time diverges, making perfectly adiabatic
evolution impossible at any finite driving rate: the state falls out of
equilibrium within a ``frozen'' interval around the transition, and
topological defects or quasiparticle excitations are left behind.  The
Kibble--Zurek (KZ) mechanism, originally introduced in the context of
cosmological phase transitions and subsequently formulated for
condensed-matter systems, relates the resulting density of excitations to
the equilibrium critical exponents and to the rate at which the transition
is crossed~\cite{Kibble76a,Zurek85a,Zurek1996}.  Its extension to quantum
phase transitions has established the KZ mechanism as a central paradigm of
nonequilibrium quantum many-body
dynamics~\cite{Zurek2005,Dziarmaga2005,Polkovnikov2005,Dziarmaga2010,
Polkovnikov2011}, recently verified on programmable quantum
processors~\cite{Kiss_2025}.

For a linear quench of duration $\tau$ across an isolated quantum critical
point, the standard KZ argument predicts a density of excitations
\begin{equation}
 n_{\mathrm{ex}} \sim \tau^{-\frac{d\nu}{1+z\nu}} ,
 \label{eq:kz_standard}
\end{equation}
where $d$ is the spatial dimension, and $\nu$ and $z$ are the
correlation-length and dynamical critical exponents, respectively.  This
scaling can be understood in terms of the freeze-out length
\begin{equation}
    \hat{\xi} \sim \tau^{\nu/(1+z\nu)} ,
    \label{eq:kz_length}
\end{equation}
which sets the characteristic length scale of the excitations generated
during passage through the critical region.  In quantum systems, the same
scaling emerges from the adiabatic-impulse picture and, for a number of
integrable models, can be related directly to a collection of Landau--Zener
transitions~\cite{Zurek2005,Dziarmaga2005}.  The KZ framework has since
been generalized to nonlinear protocols, multicritical points, extended
critical manifolds, and spatially inhomogeneous
systems~\cite{Sen2008Nonlinear,Dziarmaga2010,Polkovnikov2011}.

A particularly useful setting for exploring these questions is the Kitaev
honeycomb model~\cite{Kitaev06b}.  It describes spin-$1/2$ degrees of
freedom on a two-dimensional honeycomb lattice with bond-dependent Ising
interactions and admits an exact mapping, in each static $\mathbb{Z}_2$
gauge sector, onto free Majorana fermions.  In the thermodynamic limit, the
model possesses three gapped phases and an extended gapless phase separated
by critical lines.  Besides providing a canonical realization of a quantum
spin liquid with fractionalized excitations, its exact solvability makes it
possible to follow the real-time dynamics across phase boundaries without
relying on equilibrium or near-equilibrium approximations.

Quench dynamics in the two-dimensional Kitaev model has been investigated
extensively in the translationally invariant bulk geometry.  In particular,
Sengupta, Sen, and Mondal showed that crossing an extended critical
manifold leads to a generalization of the usual KZ
law~\cite{Sengupta08a,Mondal08a}: if a $d$-dimensional system is driven
through a critical manifold of dimension $d-m$, the excitation density
scales as
\begin{equation}
 n_{\mathrm{ex}} \sim \tau^{-\frac{m\nu}{1+z\nu}} ,
 \label{eq:kz_critical_surface}
\end{equation}
rather than according to Eq.~\eqref{eq:kz_standard}.  The Kitaev model
provides an exactly solvable two-dimensional realization of this mechanism:
the quench crosses a one-dimensional critical manifold in momentum space
($m = 1$, $\nu = z = 1$), so that the defect density follows the anomalous
power law $n_{\mathrm{ex}} \propto \tau^{-1/2}$, reflecting the codimension
of the critical manifold rather than the spatial dimensionality of the
lattice~\cite{Sengupta08a,Mondal08a}.  Subsequent studies have considered,
among other aspects, nonlinear and sudden quenches, entropy and correlation
generation, and dynamical quantum phase transitions in the Kitaev
model~\cite{Sen2008Nonlinear,Patel2012,Schmitt2015}.

Most of these analyses, however, concern the bulk system with periodic
boundary conditions in both spatial directions.  The situation becomes
qualitatively richer when one direction is opened.  Depending on the
boundary termination and coupling regime, the Kitaev model supports
Majorana modes localized near its edges~\cite{Thakurathi14a}.  Their
existence and localization are controlled by the conserved momentum
parallel to the boundary and by the relative values of $J_x$, $J_y$, and
$J_z$.  In a semi-infinite geometry, these modes approach zero energy,
whereas in a finite-width ribbon the modes localized at opposite edges
hybridize and acquire an exponentially small energy.  A finite ribbon
therefore contains both dispersive bulk modes and boundary-localized modes
whose characteristic energy and length scales depend on the system width.
This raises the question of whether KZ scaling is modified when a
two-dimensional quantum-critical system is confined in one direction, and
of how the quench populates its bulk and boundary degrees of freedom.

An important indication that boundary Majorana modes can play an active
role in KZ dynamics was provided in Ref.~\cite{ChoiMS15b}.  Considering a
one-dimensional topological superconductor, the authors generalized the KZ
construction to account for the multilevel dynamics generated by open
boundary conditions.  In contrast to a periodic system, where momentum
conservation reduces the evolution to independent few-level sectors, open
boundaries support localized Majorana zero modes and allow dynamical
transitions among several instantaneous energy levels.  The resulting
quench dynamics therefore cannot, in general, be reduced to the
conventional two-level KZ picture.  That analysis further demonstrated
that Majorana boundary states provide a distinct contribution to
quasiparticle production and introduce finite-size effects associated with
their localization and hybridization.  These results suggest that the
interplay between boundary modes and critical dynamics is an intrinsic
aspect of KZ physics in topological systems.  A natural extension is to
ask how this interplay manifests itself in a genuinely two-dimensional
system, where a conserved momentum parallel to the boundary coexists with
a finite set of transverse bulk and edge modes.

The experimental relevance of these questions has been underscored by two
recent and complementary realizations of Kitaev-honeycomb physics on
programmable quantum processors~\cite{Evered2025,Will2025}.  Using a
reconfigurable neutral-atom array, Evered \textit{et al.} implemented a
digital quantum simulation of the Kitaev honeycomb model in a cylindrical
geometry, with periodic boundary conditions along one spatial
direction~\cite{Evered2025}.  Measurement and feedforward were used to
prepare the long-range-entangled state required to encode the fermionic
statistics, whereas tunable Floquet circuits generated the subsequent
dynamics.  By measuring real-space Majorana correlations, the authors
reconstructed an effective momentum-space parent Hamiltonian and identified
the non-Abelian spin-liquid regime through an odd Chern number.  They
further demonstrated tunable fermionic dynamics and directly probed fermion
exchange statistics.  The cylindrical topology of this experiment is
closely related to the finite-width geometries considered here and shows
that the combination of bond-selective Kitaev interactions, nontrivial
boundary conditions, and Majorana observables is accessible on current
neutral-atom processors.

In a complementary superconducting-qubit implementation, Will
\textit{et al.} realized a periodically driven Kitaev model exhibiting
Floquet topological order~\cite{Will2025}.  They imaged the dynamics of
chiral edge modes, characterized emergent anyonic excitations, and
introduced an interferometric protocol for measuring a bulk dynamical
invariant associated with the transmutation of anyons.  Together, these
experiments demonstrate experimental control over both static and
intrinsically nonequilibrium aspects of two-dimensional Kitaev systems,
including topological-state preparation, emergent Majorana fermions,
physical boundaries, and highly entangled real-time evolution.

The protocols realized in Refs.~\cite{Evered2025,Will2025} are distinct
from the slow critical ramps considered in the present work.  In
particular, they do not address the universal KZ regime or the separation
of bulk-critical and boundary-generated excitations during passage through
an equilibrium quantum-critical manifold.  This distinction defines an
important missing regime for both theory and experiment.  The exact
solvability of the Kitaev model enables us to isolate universal KZ scaling
from transverse quantization, edge-mode hybridization, and other
finite-size effects, thereby providing geometry-resolved benchmarks for
future implementations of slow quenches on programmable quantum hardware.

Motivated by these theoretical and experimental developments, we
investigate how the coexistence of bulk critical modes and boundary
Majorana states modifies KZ dynamics when the problem is extended from a
one-dimensional topological system to a two-dimensional quantum spin
liquid.  We consider cylindrical honeycomb geometries that are periodic
along the direction parallel to the boundary and finite in the transverse
direction, with two parallel boundaries of either zigzag or armchair type,
together with the corresponding fully periodic tubes as boundary-free
references.  In the flux-free sector, the spin Hamiltonian maps exactly
onto a quadratic problem of noninteracting Majorana fermions.  Translation
invariance along the periodic direction allows the dynamics to be
decomposed into independent momentum sectors, whereas the open transverse
direction retains the finite-width structure of the bulk subbands and
boundary-localized states.  The bipartite structure of the honeycomb
lattice gives rise to a chiral symmetry that enforces a spectrum symmetric
about zero energy and provides a natural framework for identifying and
characterizing the Majorana edge modes.

We drive the system across its quantum-critical region by linearly ramping
the coupling $J_z$ on the $z$-links while keeping $J_x$ and $J_y$ fixed.
Because the $\mathbb{Z}_2$ flux sector is conserved by this protocol, the
Majorana Hamiltonian remains quadratic throughout the evolution.  Starting
from the initial instantaneous many-body ground state, the complete
time-dependent problem can therefore be solved exactly in each conserved
momentum sector.  We quantify the nonadiabatic response by projecting the
time-evolved occupied subspace onto the instantaneous positive-energy
subspace and use the resulting quasiparticle density to investigate the KZ
regime.

Entanglement provides a complementary, basis-independent diagnostic of the
quench~\cite{Calabrese05a,Cincio07a,Alba17a}.  For Gaussian fermionic
states both the entanglement entropy and the entanglement spectrum are
accessible from two-point functions~\cite{Peschel03a,Peschel2009}, the
latter reflecting the topology of the state through its edge-state
structure~\cite{LiHaldane2008,Fidkowski2010,YaoQi2010,QiKatsuraLudwig2012},
and the fermionic logarithmic negativity distinguishes quantum from
classical correlations in the post-quench state~\cite{Shapourian17a}.  We
therefore track the quench-generated entanglement entropy and negativity
across bipartitions of both zigzag and armchair orientation, alongside the
defect counts.

Our results reveal a pronounced interplay among KZ physics, transverse
confinement, and Majorana edge modes.  The excitation density exhibits a
width-dependent crossover governed by the competition between the KZ
freeze-out length and the transverse system size $N_2$, naturally
organized by the scaling variable
\begin{equation}
    \frac{N_2}{\hat{\xi}} \sim \frac{N_2}{\sqrt{\tau}} .
    \label{eq:finite_width_kz_intro}
\end{equation}
Three results emerge.  (i)~On the tubes the defect density per unit length
collapses onto a single function of $N_2/\sqrt{\tau}$ and is independent of
the wrapping direction.  (ii)~The zigzag ribbon follows the same scaling
function but is cut off exponentially at $N_2/\sqrt{\tau} \lesssim 20$: the
open boundary converts the critical gap closing into the formation of the
Majorana edge doublet, so that the spectral gap of the transverse chain
never closes at finite width; the occupation of the edge doublet is in
turn pinned to exactly one fermion by inversion parity.  (iii)~The
armchair ribbon evades
the cutoff altogether: inversion-parity-protected zero-energy crossings at
zero longitudinal momentum seed a channel of transient Majorana edge modes
with an exponentially small
gap at every momentum, which acts as an exact level crossing for any
subexponential ramp and deposits a width-independent defect number set by
the even--odd parity of the width---exactly one per unit length for odd
widths, a transient plateau near one half for even widths.  Underlying both departures from the tube behavior is
a single physical origin: the zero-energy edge modes required by the
bulk--boundary correspondence, which remove the critical gap closing from
the zigzag bulk and supply the armchair ribbon with its
adiabaticity-immune channel.  The behavior thus changes qualitatively depending on whether
the boundary channels are included in the excitation count or only the
bulk bands are retained, providing a direct way of distinguishing
boundary-generated excitations from the conventional bulk KZ contribution.
The quench-generated entanglement entropy and logarithmic negativity,
across both inequivalent cut orientations, inherit this entire structure.

More broadly, our analysis demonstrates that boundaries need not constitute
a passive finite-size correction to KZ dynamics.  In topological and
fractionalized systems, the bulk--boundary correspondence equips the edges
with zero-energy modes whose characteristic scales are parametrically
different from those of the bulk and which consequently participate
directly in the breakdown of adiabaticity.  The Kitaev ribbon
provides a particularly transparent realization of this situation because
bulk and edge contributions can be resolved at the single-particle level
within an exactly solvable many-body model.  Our results connect universal
slow-quench dynamics with the boundary Majorana physics of quantum spin
liquids and establish a controlled benchmark for studying confined
topological matter on emerging quantum-simulation platforms.

The remainder of this paper is organized as follows.
Section~\ref{sec:ribbons} reviews the model, the ribbon geometries, and the
static edge-mode phase diagrams.  Section~\ref{sec:quench} specifies the
quench protocol, the observables, and the numerical method.
Sections~\ref{sec:defects} and~\ref{sec:entanglement} present the results
for defect production and entanglement generation, respectively, and
Sec.~\ref{sec:summary} concludes.

\section{Kitaev Ribbons}
\label{sec:ribbons}

Here we first review the static properties of the Kitaev ribbons, focusing
on the ground-state phase diagram and on the zero-energy edge modes with
their edge-mode phase diagrams.  We then identify two distinct notions of
parity which, together with the zero-energy edge modes, turn out to shape
the KZ scaling analyzed in Secs.~\ref{sec:defects}
and~\ref{sec:entanglement}.  Throughout we follow the conventions of
Ref.~\cite{Kitaev06b}.

\subsection{Kitaev Model}
\label{sec:model}

\begin{figure}
  \centering
  \begin{overpic}[width=0.98\linewidth]{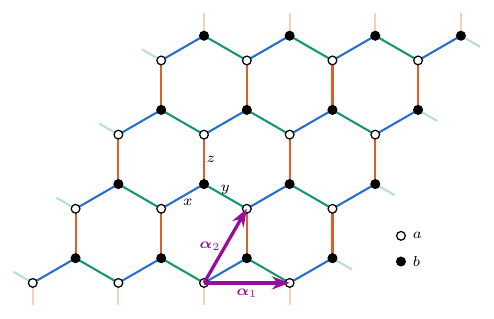}
    \put(0,47){(a)}
  \end{overpic}\\[6pt]
  \begin{overpic}[width=0.72\linewidth]{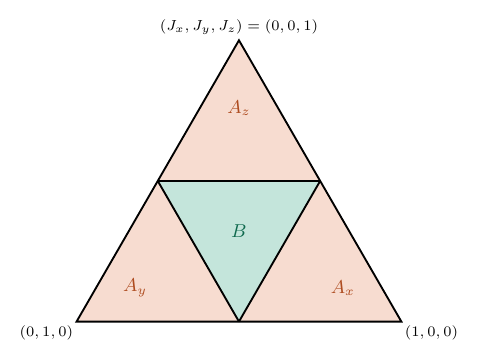}
    \put(2,64){(b)}
  \end{overpic}
  \caption{(a)~Honeycomb lattice with the primitive vectors $\bfalpha_1$ and
    $\bfalpha_2$.  Open (filled) circles denote the $a$ ($b$) sublattice;
    blue, green, and orange bonds denote the $x$-, $y$-, and $z$-links,
    respectively.  (b)~Ground-state phase diagram of the Kitaev model on the
    plane $J_x + J_y + J_z = 1$ ($J_\alpha \geq 0$).  The three gapped phases
    $A_x$, $A_y$, and $A_z$ surround the gapless phase $B$ defined by the
    triangle inequalities~\eqref{eq:triangle}.}
  \label{fig:lattice}
\end{figure}

The Kitaev model describes spin-$1/2$ degrees of freedom on the honeycomb
lattice with strongly anisotropic, bond-dependent Ising
interactions~\cite{Kitaev06b},
\begin{equation}
  \hat{H} = -\sum_{\alpha \in \{x,y,z\}} \;
  \sum_{\langle ij \rangle_\alpha} J_\alpha\,
  \hat\sigma_i^\alpha \hat\sigma_j^\alpha ,
  \label{eq:spinH}
\end{equation}
where $\langle ij \rangle_\alpha$ denotes the nearest-neighbor pairs
connected by an $\alpha$-link and $J_\alpha$ are the corresponding exchange
couplings.  The honeycomb lattice consists of two triangular sublattices,
$a$ and $b$; we span it by the primitive vectors $\bfalpha_1 = (1,0)$ and
$\bfalpha_2 = (\tfrac{1}{2}, \tfrac{\sqrt{3}}{2})$ (lattice constant set to
unity) and label the two sites of the unit cell at position
$\ell_1\bfalpha_1 + \ell_2\bfalpha_2$ by $a_{\ell_1,\ell_2}$ and $b_{\ell_1,\ell_2}$.  
The three
link types are then $a_{\ell_1,\ell_2}\text{--}b_{\ell_1,\ell_2}$
($x$-links), $b_{\ell_1,\ell_2}\text{--}a_{\ell_1+1,\ell_2}$ ($y$-links),
and $b_{\ell_1,\ell_2}\text{--}a_{\ell_1,\ell_2+1}$ ($z$-links), the
$z$-links being vertical; see Fig.~\ref{fig:lattice}(a).

Following Kitaev~\cite{Kitaev06b}, each spin is represented by four Majorana
fermions $\hat d_j^x$, $\hat d_j^y$, $\hat d_j^z$, and $\hat c_j$ through
\begin{equation}
  \hat\sigma_j^\alpha = \ii\, \hat d_j^\alpha \hat c_j ,
  \label{eq:majrep}
\end{equation}
subject to the local constraint
$\hat d_j^x \hat d_j^y \hat d_j^z \hat c_j = 1$.  In this representation the
bond operators $\hat u_{ij} = \ii\, \hat d_i^\alpha \hat d_j^\alpha$ commute
with the Hamiltonian and with one another, and reduce
Eq.~\eqref{eq:spinH} to a bilinear form in the itinerant Majorana operators
$\hat c_j$ within each sector of the static $\mathbb{Z}_2$ gauge field
$\{u_{ij} = \pm 1\}$.  The physical spectrum depends on the gauge field only
through the fluxes on the hexagonal plaquettes, and the ground state resides
in the flux-free sector.  Fixing the gauge $u_{ij} = +1$ on every link
oriented from the $b$ to the $a$ sublattice, and writing
$\hat a_{\ell_1,\ell_2}$ and $\hat b_{\ell_1,\ell_2}$ for the itinerant Majorana operators
on the two sublattices (instead of the sublattice-agnostic notation $\hat c_j$),
the Hamiltonian in the flux-free sector takes the quadratic form
\begin{multline}
  \hat{H} = \ii \sum_{\ell_1,\ell_2}
  \bigl( J_x\, \hat a_{\ell_1,\ell_2} \hat b_{\ell_1,\ell_2}
       + J_y\, \hat a_{\ell_1+1,\ell_2} \hat b_{\ell_1,\ell_2} \\
       + J_z\, \hat a_{\ell_1,\ell_2+1} \hat b_{\ell_1,\ell_2} \bigr) .
  \label{eq:majH}
\end{multline}

The ground-state phase diagram in the plane $J_x + J_y + J_z = \text{const}$
is the well-known triangle shown in Fig.~\ref{fig:lattice}(b).  The spectrum
of Eq.~\eqref{eq:majH} on the torus is gapless if and only if the three
couplings satisfy the triangle inequalities
\begin{equation}
  |J_x| \leq |J_y| + |J_z| , \quad
  |J_y| \leq |J_z| + |J_x| , \quad
  |J_z| \leq |J_x| + |J_y| ,
  \label{eq:triangle}
\end{equation}
which define the gapless phase $B$ with two Dirac points; when one coupling
dominates, the spectrum is gapped and the model is in one of the three
equivalent phases $A_x$, $A_y$, or $A_z$, each mapping onto a toric-code-type
$\mathbb{Z}_2$ topological phase~\cite{Kitaev06b}.

\subsection{Zigzag and Armchair Ribbons}
\label{sec:geometry}

\begin{figure*}
  \centering
  \raisebox{9mm}{\includegraphics[width=0.46\linewidth]{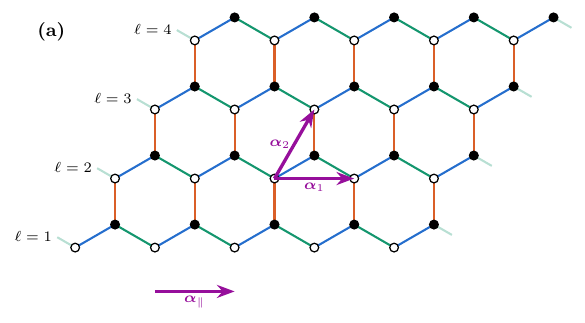}}
  \hfill
  \includegraphics[width=0.235\linewidth]{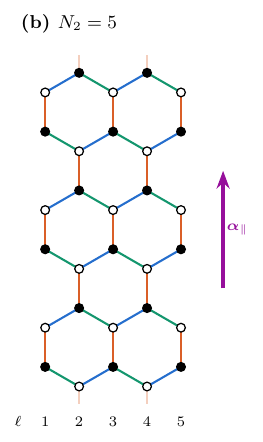}
  \hfill
  \includegraphics[width=0.27\linewidth]{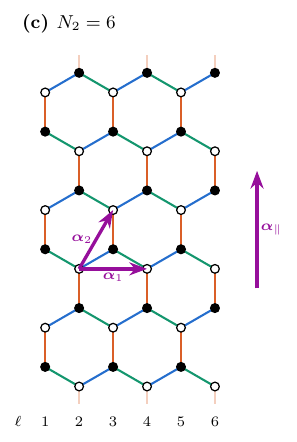}
  \caption{Ribbon geometries.  (a)~Zigzag ribbon: periodic along its
    longitudinal direction $\bfalpha_\parallel = \bfalpha_1$ with $N_x$ unit
    cells and open across, comprising $N_2$ zigzag chains labeled
    $\ell = 1, \dots, N_2$; the lower edge terminates on the $a$ sublattice
    and the upper edge on the $b$ sublattice.  (b), (c)~Armchair ribbons of
    width $N_2 = 5$ and $N_2 = 6$: periodic along
    $\bfalpha_\parallel = 2\bfalpha_2 - \bfalpha_1$ with $N_x$ two-site
    cells per column and open across, comprising $N_2$ vertical dimer lines
    labeled $\ell = 1, \dots, N_2$.  The primitive lattice vectors
    $\bfalpha_1$ and $\bfalpha_2$ are indicated in panels (a) and (c);
    faded bonds indicate the periodic continuation.}
  \label{fig:ribbons}
\end{figure*}

We consider ribbons of the two inequivalent orientations shown in
Fig.~\ref{fig:ribbons}.  The \emph{zigzag ribbon}
[Fig.~\ref{fig:ribbons}(a)] is periodic along its longitudinal direction
$\bfalpha_\parallel = \bfalpha_1$, with $N_x$ unit cells, and open in the
transverse direction, comprising $N_2$ zigzag chains labeled by
$\ell = 1, \dots, N_2$; the $z$-links connect consecutive chains, so
that the two edges terminate on opposite sublattices.  The \emph{armchair
ribbon} [Figs.~\ref{fig:ribbons}(b) and \ref{fig:ribbons}(c)] is periodic
along its longitudinal direction
$\bfalpha_\parallel = 2\bfalpha_2 - \bfalpha_1 = (0,\sqrt3)$ and open
across, comprising $N_2$ vertical lines of $z$-dimers, likewise labeled by
$\ell = 1, \dots, N_2$; in contrast
to the zigzag case, both edges cut through $x$- and $y$-links and preserve
the $z$-dimers, and the parity of $N_2$ plays an essential role below.
Closing either ribbon in the transverse direction yields a torus; we refer
to it as a \emph{tube} wrapped in the zigzag or armchair direction,
respectively, and use it as the edge-free reference geometry.

Owing to the translational invariance along the ribbon, the Majorana
operators are Fourier transformed with respect to the longitudinal cell
index.  For the zigzag ribbon we abbreviate the longitudinal and
transverse cell indices by $x \equiv \ell_1$ and $\ell \equiv \ell_2$,
respectively, and define
\begin{equation}
  \hat a_\ell(k) = \frac{1}{\sqrt{2N_x}} \sum_{x} \ee^{-\ii k x}\,
  \hat a_{x,\ell} ,
  \label{eq:FT}
\end{equation}
with $k = 2\pi p/N_x$ ($p \in \mathbb{Z}$), and analogously for
$\hat b_\ell(k)$; the armchair ribbon is treated identically, with the
cell index along $\bfalpha_\parallel$ in place of $x$ and the dimer-line
index in place of $\ell$.  Since the Majorana
operators are Hermitian, the Fourier modes obey the reality constraint
$\hat a_\ell^\dagger(k) = \hat a_\ell(-k)$, so that
only half of the Brillouin zone, $0 < k < \pi$, labels independent complex
fermion modes.  
For convenience, but without affecting the results, we exclude the two self-conjugate momenta $k = 0$ and $k = \pi$ (for $N_x$ even): there the constraint reads
$\hat a_\ell^\dagger(k) = \hat a_\ell(k)$, so the corresponding Fourier
modes remain Hermitian---genuine Majorana rather than complex
fermions---and the $2N_2$ of them recombine among themselves into only
$N_2$ complex fermions.  
These two sectors contribute to all intensive
quantities at relative order $1/N_x$ and are omitted throughout.
Up to this correction, in terms of the spinor
$\hat\Psi(k) = \bigl( \hat a_1(k), \dots, \hat a_{N_2}(k),
\hat b_1(k), \dots, \hat b_{N_2}(k) \bigr)^{\mathsf T}$,
the Hamiltonian~\eqref{eq:majH} decomposes as
\begin{equation}
  \hat{H} = 2 \sum_{0<k<\pi} \hat\Psi^\dagger(k)\, H(k)\, \hat\Psi(k) ,
  \label{eq:Hsum}
\end{equation}
where the Bloch Hamiltonian has the chiral block form
\begin{equation}
  H(k) = \begin{pmatrix} 0 & \ii D(k) \\ -\ii D^\dagger(k) & 0 \end{pmatrix} .
  \label{eq:Hblock}
\end{equation}
For the zigzag ribbon the $N_2 \times N_2$ block is lower bidiagonal,
\begin{equation}
  D_{\mathrm{zz}}(k)
  = \bigl( J_x + J_y\, \ee^{-\ii k} \bigr) \mathbbm{1}_{N_2} + J_z L ,
  \label{eq:Dzz}
\end{equation}
where $L_{\ell\ell'} = \delta_{\ell,\ell'+1}$.  For the armchair ribbon the
block is tridiagonal,
\begin{equation}
  D_{\mathrm{ac}}(k) = J_z\, \mathbbm{1}_{N_2}
   + J_x L^{\mathsf T} + J_y\, \ee^{\ii k} L .
  \label{eq:Dac}
\end{equation}
In either case the eigenstates $\ket{\psi_n^\pm(k)}$ of $H(k)$ have
energies $\varepsilon_n^\pm(k) = \pm 2 s_n(k)$, $n = 0, 1, \dots, N_2-1$,
where $s_0(k) \le s_1(k) \le \cdots \le s_{N_2-1}(k)$ are the singular
values of $D(k)$ in ascending order.

At fixed $k$, each block $D(k)$ is a banded Toeplitz matrix, and the
topological character of the corresponding transverse chain is encoded in
its \emph{Bloch symbol}: the scalar function obtained by closing the chain
periodically and replacing the shift operator $L$ by its Bloch eigenvalue
$\ee^{\ii q}$, with $q$ the momentum conjugate to the transverse index
$\ell$.  For the two ribbons,
\begin{equation}
  F^{\mathrm{zz}}_k(q) = \bigl( J_x + J_y\, \ee^{-\ii k} \bigr)
  + J_z\, \ee^{\ii q}
  \label{eq:Fzz}
\end{equation}
and
\begin{equation}
  F^{\mathrm{ac}}_k(q) = J_z + J_x\, \ee^{-\ii q} + J_y\, \ee^{\ii (k+q)} .
  \label{eq:Fac}
\end{equation}
As $q$ sweeps the transverse Brillouin zone, the symbol traces a closed
curve in the complex plane---a circle of radius $|J_z|$ centered at
$J_x + J_y \ee^{-\ii k}$ for the zigzag ribbon, an ellipse centered at
$J_z$ for the armchair ribbon---and we call the symbol
\emph{topologically wound} when this curve encircles the origin, that is,
when its winding number is nonzero.  A nonzero winding is the topological
criterion of the chiral symmetry class: it forces a singular value of the
open chain that vanishes in the semi-infinite limit---one Majorana zero
mode exponentially localized at each edge---and that survives at finite
width with a magnitude exponentially small in $N_2$.  The winding can
change only where the symbol passes through the origin, so the boundaries
of the edge-mode regions are precisely the momenta at which the gap of
the infinitely wide transverse chain closes; in the gapless phase $B$
these are the projections of the two bulk Dirac points onto the $k$ axis.

\begin{figure}
  \centering
  \includegraphics[width=0.48\linewidth]{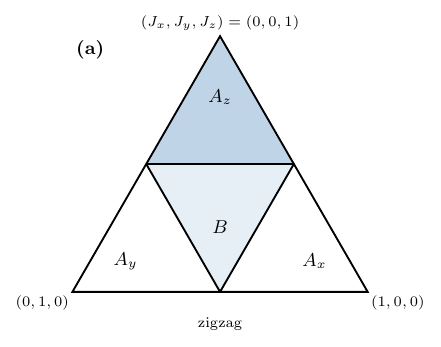}
  \hfill
  \includegraphics[width=0.48\linewidth]{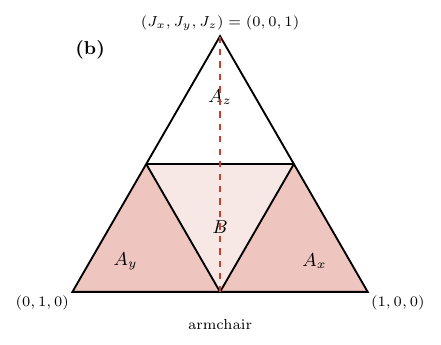}\\[8pt]
  \includegraphics[width=0.98\linewidth]{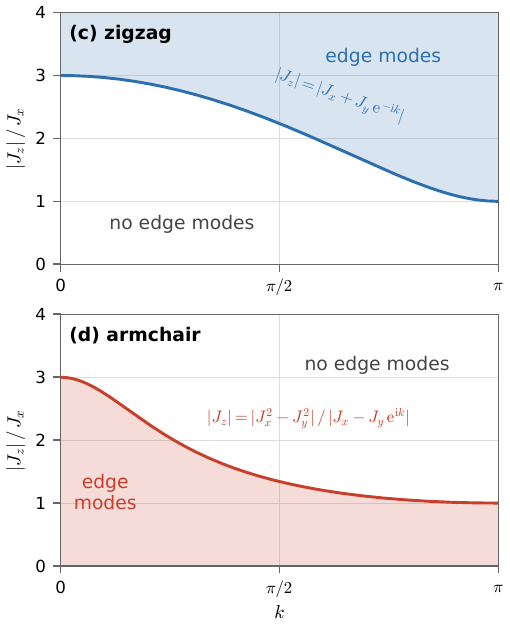}
  \caption{Edge-mode phase diagrams.  (a), (b)~Schematic representation on
    the coupling triangle $J_x + J_y + J_z = 1$ ($J_\alpha \geq 0$) for the
    zigzag and armchair ribbons, respectively: dark shading marks the
    couplings at which edge modes exist for every momentum $k$, light
    shading those at which they occupy a finite window of $k$.  Zigzag edge
    modes exist for $|J_z| > |J_x - J_y|$, i.e., throughout $B$ and $A_z$;
    armchair edge modes exist for $|J_z| < J_x + J_y$, i.e., throughout
    $A_x$, $A_y$, and $B$, except on the line $J_x = J_y$ (dashed), on
    which the window closes.  (c), (d)~The same conditions,
    Eqs.~\eqref{eq:zzedge} and~\eqref{eq:acedge}, resolved in the
    $(k, |J_z|)$ plane at $J_x = 1$ and $J_y = 2$, the couplings used
    throughout this work; the boundaries are the gap closings of the
    transverse chains.  The two regions are complementary: a $J_z$ ramp
    terminates deep in the zigzag edge phase but merely traverses the
    armchair edge window.}
  \label{fig:edgephases}
\end{figure}

For the zigzag ribbon, Eq.~\eqref{eq:Dzz} is the chiral block of an open
Su--Schrieffer--Heeger chain with intracell hopping
$|J_x + J_y \ee^{-\ii k}|$ and intercell hopping $|J_z|$, and the circle
traced by $F^{\mathrm{zz}}_k$ encircles the origin precisely when
\begin{equation}
  \bigl| J_x + J_y\, \ee^{-\ii k} \bigr| < |J_z| \qquad \text{(zigzag)} ,
  \label{eq:zzedge}
\end{equation}
which reproduces the known zigzag edge-mode phase
diagram~\cite{Thakurathi14a}: at fixed couplings the edge modes occupy the
momentum interval around $k = \pi$ where Eq.~\eqref{eq:zzedge} holds,
Figs.~\ref{fig:edgephases}(a) and \ref{fig:edgephases}(c).  For the
armchair ribbon, the ellipse traced by $F^{\mathrm{ac}}_k$ encircles the
origin precisely when
\begin{equation}
  |J_z|\, \bigl| J_x - J_y \ee^{\ii k} \bigr| < \bigl| J_x^2 - J_y^2 \bigr|
  \qquad \text{(armchair)} ,
  \label{eq:acedge}
\end{equation}
a window centered at $k = 0$,
Figs.~\ref{fig:edgephases}(b) and \ref{fig:edgephases}(d).  The two
conditions are complementary: for fixed $J_x \neq J_y$ the zigzag edge
modes exist at large $|J_z|$ and the armchair edge modes at small $|J_z|$,
and on the isotropic line $J_x = J_y$ the armchair window closes entirely.

At fixed $k$, the edge modes, when present, form the doublet
$\ket{\psi_0^\pm(k)}$ with energies $\varepsilon_0^\pm(k)$ exponentially
small in the width $N_2$, while the remaining eigenstates
$\ket{\psi_n^\pm(k)}$ with $n \geq 1$ constitute the bulk bands.  For the
finite-width analysis below it is convenient to decompose the
single-particle space of each momentum sector accordingly, into the
positive band, the negative band, and the edge doublet, with the
orthogonal projectors
\begin{subequations}
\label{eq:projectors}
\begin{align}
  P_\pm(k) &:= \sum_{n=1}^{N_2-1} \ket{\psi_n^\pm(k)}\bra{\psi_n^\pm(k)} ,
  \label{eq:Pbulk} \\
  P_0(k) &:= \sum_{\pm} \ket{\psi_0^\pm(k)}\bra{\psi_0^\pm(k)} .
  \label{eq:Pedge}
\end{align}
\end{subequations}
Whether a sector supports edge modes is decided by the winding criterion
above; when it does not, the lowest doublet is a bulk state like any
other, and we set $P_0(k) = 0$ and include $n = 0$ in the band
projectors, $P_\pm(k) = \sum_{n=0}^{N_2-1}
\ket{\psi_n^\pm(k)}\bra{\psi_n^\pm(k)}$.  When present, the edge sector is
identified spectrally, as the singular subspace of $D(k)$ belonging to the
smallest singular value $s_0(k)$, which remains well conditioned even
where the doublet splitting underflows machine precision.

Beyond the zero-energy edge modes, we identify two distinct notions of
parity that organize the low-energy structure of the ribbons and, as shown
in Secs.~\ref{sec:defects} and~\ref{sec:entanglement}, control their quench
dynamics, each decisive for one termination: the \emph{inversion parity}
of the single-particle states, which governs the zigzag ribbon, and the
even--odd parity of the width, which governs the armchair ribbon.  We
begin with the former.  For $J_x \neq J_y$ the only
point-group operation of either ribbon is the in-plane $C_2$ rotation
about its center, which reverses the longitudinal direction, maps the
transverse index $\ell \to N_2 + 1 - \ell$, and---since the two edges
terminate on opposite sublattices---exchanges $a$ and $b$.  In the block
form of Eq.~\eqref{eq:Hblock} it is represented by
\begin{equation}
  P = \begin{pmatrix} 0 & -\ii R \\ \ii R & 0 \end{pmatrix} ,
  \label{eq:P}
\end{equation}
where $R_{\ell\ell'} = \delta_{\ell, N_2+1-\ell'}$ reflects the transverse
index and $P^2 = \mathbbm{1}$.  Because the reversal of the longitudinal
direction sends $k \to -k$, $P$ does not commute with the Bloch
Hamiltonian as written; using $R L R = L^{\mathsf T}$ one finds instead
\begin{equation}
  P\, H(k)\, P = H(-k) ,
  \label{eq:Pobstruction}
\end{equation}
so that inversion parity appears to be a good quantum number only at the
self-conjugate momenta $k = 0$ and $k = \pi$.  On the zigzag ribbon this
obstruction is removable.  Writing
$J_x + J_y \ee^{-\ii k} = |J_x + J_y \ee^{-\ii k}|\, \ee^{\ii \varphi_k}$,
the phase $\varphi_k$ is odd in $k$, and the $k$-dependent gauge
transformation $\hat a_\ell(k) \to \ee^{\ii \ell \varphi_k} \hat a_\ell(k)$,
$\hat b_\ell(k) \to \ee^{\ii (\ell+1) \varphi_k} \hat b_\ell(k)$---which
preserves the Majorana reality constraint precisely because
$\varphi_{-k} = -\varphi_k$---renders the chiral block real,
\begin{equation}
  \tilde D_{\mathrm{zz}}(k)
  = \bigl| J_x + J_y\, \ee^{-\ii k} \bigr|\, \mathbbm{1}_{N_2} + J_z L .
  \label{eq:Dtilde}
\end{equation}
In this gauge $R \tilde D_{\mathrm{zz}} R = \tilde D_{\mathrm{zz}}^{\mathsf T}$,
so that $P$ commutes with the transformed Bloch Hamiltonian $\tilde H(k)$
at \emph{every} momentum, and the unitary $V$ whose columns are the parity
eigenvectors $(u, \pm \ii R u)/\sqrt{2}$ block-diagonalizes it,
\begin{equation}
  V^\dagger\, \tilde H(k)\, V
  = \begin{pmatrix} -S(k) & 0 \\ 0 & S(k) \end{pmatrix} ,
  \label{eq:parityblocks}
\end{equation}
with $S(k) := \tilde D_{\mathrm{zz}}(k)\, R$ a real symmetric
$N_2 \times N_2$ matrix whose eigenvalues are, up to sign, the singular
values $s_n(k)$.  The $2N_2$-dimensional problem thus collapses to a single
real symmetric matrix, and since inversion exchanges the sublattices, the
chiral symmetry anticommutes with $P$ and maps one parity sector onto the
other: every chiral pair $\pm 2 s_n(k)$ is a \emph{parity doublet}, with
its two members residing in opposite sectors.  This applies in particular
to the edge doublet $\ket{\psi_0^\pm(k)}$, whose members are the even and
odd combinations of the two edge-localized Majorana modes and whose
exponentially small splitting is a parity splitting.  Because the ramp of
$J_z$ changes only $S(k)$ and leaves the block structure intact, the two
parity sectors evolve independently during the quench, a fact exploited
in Secs.~\ref{sec:quench} and \ref{sec:defects}.  On the armchair ribbon
the same $C_2$ operation and the same obstruction~\eqref{eq:Pobstruction}
apply, but the phase $\ee^{\ii k}$ on the $J_y$ entries of
Eq.~\eqref{eq:Dac} cannot be removed by any site-local gauge
transformation unless $\ee^{2\ii k} = 1$; there, inversion parity is a
good quantum number only at $k = 0$ and $k = \pi$.

The second notion, the \emph{even--odd} parity of the ribbon width $N_2$
(the \emph{width parity} for short), is peculiar to the armchair ribbon,
where inversion parity itself survives only at the two self-conjugate
momenta.  Although the self-conjugate momentum $k = 0$ is
excluded from our mode decomposition~\eqref{eq:Hsum}, the structure of $D_{\mathrm{ac}}(k)$ of Eq.~\eqref{eq:Dac} at $k = 0$ serves as a
reference for the finite momenta.  At $k = 0$ the phase factor in
Eq.~\eqref{eq:Dac} disappears, the transverse chain becomes real, and
inversion parity is restored, decomposing $H(0)$ into two sectors as in
Eq.~\eqref{eq:parityblocks}, so that levels of opposite inversion parity
may cross at zero energy.  The spectrum of the tridiagonal Toeplitz block
is classical~\cite{Losonczi92a}, and it shows that exactly $N_2$ such
inversion-parity-protected zero-energy crossings occur, at the discrete
couplings
\begin{equation}
  J_z = 2\sqrt{J_x J_y}\, \cos\frac{\pi p}{N_2 + 1} ,
  \qquad p = 1, \dots, N_2 .
  \label{eq:crossings}
\end{equation}

At the finite momenta $0 < k < \pi$ the phase factor $\ee^{\ii k}$
cannot be removed, the inversion protection is lost, and the
crossings~\eqref{eq:crossings} generically open into avoided crossings
with gaps proportional to $J_y \sin k$. One exception survives, decided by the width parity.  
At $J_z = 0$ the block $D_{\mathrm{ac}}(k)$
is a tridiagonal matrix with vanishing diagonal, whose determinant
vanishes identically for odd dimension; for odd $N_2$ one singular value
is therefore exactly zero at $J_z = 0$ for \emph{every} momentum, and the
corresponding zero-energy crossing---the one that reduces to the central
member, $p = (N_2+1)/2$, of Eq.~\eqref{eq:crossings} at $k = 0$---remains
exact throughout the Brillouin zone.  For even $N_2$ no such channel
exists, and every crossing acquires a finite gap away from $k = 0$.  The
width parity, rather than the magnitude of $N_2$, thereby controls the
low-energy structure at every finite momentum, a fact that underlies the
defect plateaus of Sec.~\ref{sec:defects}.

\section{Quench Process}
\label{sec:quench}

We drive the system by a linear ramp of the $z$-bond coupling,
\begin{equation}
  J_z(t) = J_0\, \frac{t}{\tau} , \qquad -T \leq t \leq T ,
  \label{eq:ramp}
\end{equation}
at fixed $J_x = J_0$ and $J_y = 2J_0$ ($J_0 \equiv 1$ sets the units), with
$J_{\max} = J_0 T/\tau$ chosen deep in the gapped phases so that both endpoints
of the protocol are adiabatic; we use $J_{\max} = 6\,J_0$ throughout,
having checked that larger values leave the results unchanged.  The system is prepared in the many-body
ground state at $t = -T$; on the zigzag ribbon, where the initial edge
doublet is split only by an energy exponentially small in $N_2$, the
occupied member is identified exactly through the inversion-parity
decomposition~\eqref{eq:parityblocks} rather than through the sign of a
numerically underflowing eigenvalue.  Because the momentum $k$ along the ribbon remains a good
quantum number throughout the ramp, the dynamics factorizes into independent
$2N_2$-level Landau--Zener problems~\cite{Zener32a}, one for each
$0 < k < \pi$.

\begin{figure}
  \centering
  \includegraphics[width=0.99\linewidth]{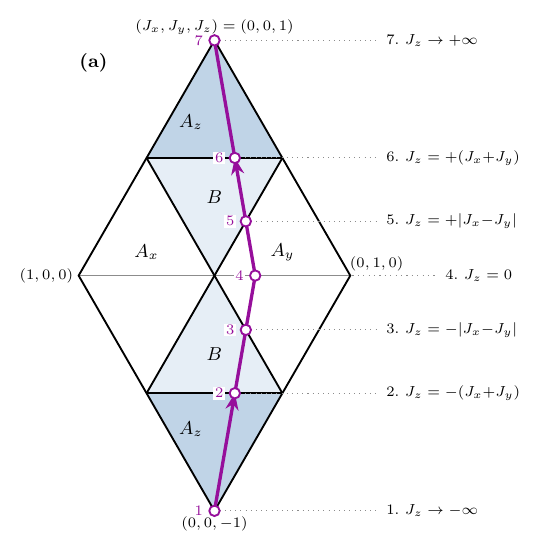}\\[6pt]
  \includegraphics[width=0.99\linewidth]{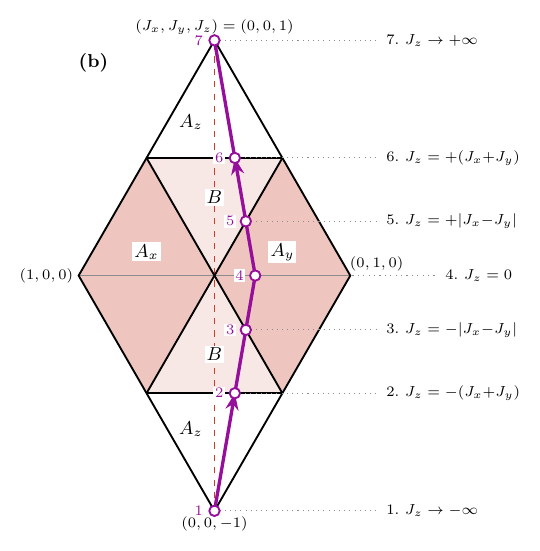}
  \caption{Trajectory of the quench on the phase diagram extended by the
    sign of $J_z$: the upper and lower triangles are the phase diagrams for
    $J_z > 0$ and $J_z < 0$, respectively, on the plane
    $|J_x| + |J_y| + |J_z| = 1$.  The arrowed line is the
    ramp~\eqref{eq:ramp} at the fixed ratio $J_x : J_y = 1 : 2$; the labeled
    points mark the endpoints deep in the gapped phases ($1$, $7$), the
    crossings of the gapless-phase boundaries at $J_z = \mp(J_x + J_y)$
    ($2$, $6$) and $J_z = \mp|J_x - J_y|$ ($3$, $5$), and the midpoint
    $J_z = 0$ ($4$).  Shading follows Figs.~\ref{fig:edgephases}(a)
    and~\ref{fig:edgephases}(b): (a)~zigzag and (b)~armchair edge-mode
    regions, with dark shading where edge modes exist at every momentum,
    light shading where they occupy a finite momentum window, and the
    dashed line marking $J_x = J_y$.}
  \label{fig:trajectory}
\end{figure}

Figure~\ref{fig:trajectory} locates the protocol on the phase diagram,
extended by the sign of $J_z$.  At the fixed ratio $J_x : J_y = 1 : 2$ the
ramp runs along the straight line connecting the two $A_z$ vertices and
crosses the gapless phase $B$ twice, at
$-(J_x + J_y) \leq J_z \leq -|J_x - J_y|$ (points $2$--$3$) and at
$|J_x - J_y| \leq J_z \leq J_x + J_y$ (points $5$--$6$), passing through
the gapped $A_y$ phase in between (points $3$--$5$).  Relative to the
edge-mode regions of Fig.~\ref{fig:edgephases}, the two terminations
experience the ramp in opposite ways: the zigzag edge region,
$|J_z| > |J_x - J_y|$ [shaded in Fig.~\ref{fig:trajectory}(a)], contains
both the initial and the final segments of the trajectory, so the zigzag
edge doublet is present at both endpoints of the protocol; the armchair
edge region, $|J_z| < J_x + J_y$ [shaded in
Fig.~\ref{fig:trajectory}(b)], is merely traversed between points $2$
and~$6$, so the armchair edge modes exist only transiently during the
ramp.

\begin{figure}
  \centering
  \includegraphics[width=0.92\linewidth]{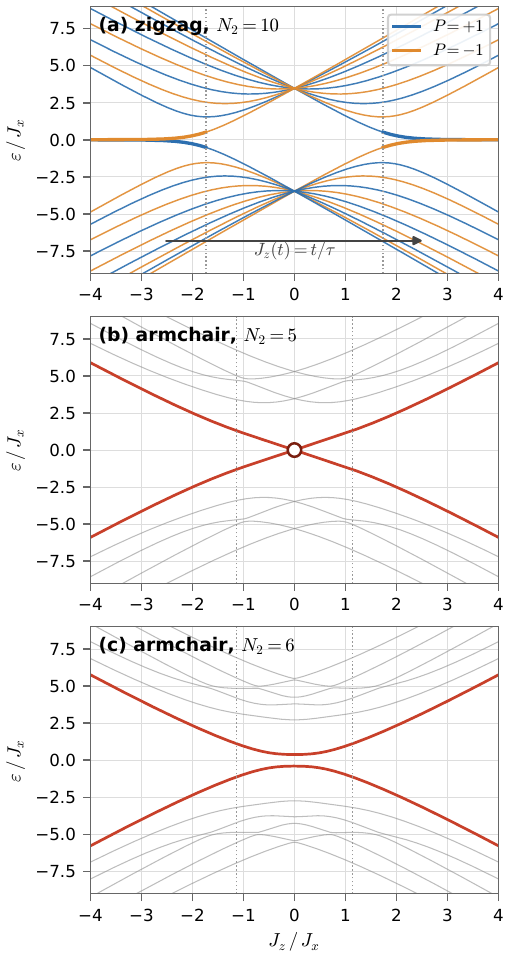}
  \caption{Single-particle spectral flow across the ramp at the
    representative momentum $k = 2\pi/3$.  (a)~Zigzag ribbon
    ($N_2 = 10$), resolved by inversion parity according to
    Eq.~\eqref{eq:parityblocks}: levels with $P = +1$ (blue) and $P = -1$
    (orange) cross freely, whereas levels of equal parity repel.  The
    transverse chain closes its gap at
    $|J_z| = |J_x + J_y \ee^{-\ii k}| = \sqrt{3}$, marked by the dotted
    lines, beyond which the edge doublet (thick lines) approaches zero
    energy exponentially; its two members carry opposite parity.  At
    finite width the minimum gap remains finite at every $k$.  (b), (c)~Armchair ribbons of odd and even width,
    $N_2 = 5$ and $N_2 = 6$, corresponding to Figs.~\ref{fig:ribbons}(b)
    and \ref{fig:ribbons}(c); the small widths are chosen so that the
    low-energy gaps are resolvable on this scale.  Inside the transient
    edge window of Eq.~\eqref{eq:acedge}, whose boundary is marked by the
    dotted lines, the lowest doublet (red) is the topologically wound edge
    branch, whose gap is exponentially small in the width.  For odd $N_2$
    [panel~(b)] the doublet crosses zero energy exactly at $J_z = 0$
    (circle), at every momentum; for even $N_2$ [panel~(c)] the crossing
    is avoided and the minimum gap remains finite.  Gray lines show the
    remaining levels.}
  \label{fig:spectralflow}
\end{figure}

Figure~\ref{fig:spectralflow} shows the single-particle spectral flow of the
two ribbons.  On the zigzag ribbon the transverse chain at momentum $k$
passes through avoided gap closings at $|J_z| = |J_x + J_y \ee^{-\ii k}|$;
at finite width
the minimum gap is bounded below by an algebraic floor (see
Sec.~\ref{sec:defects}), so that every momentum sector eventually becomes
adiabatic at large $\tau$.  The parity resolution of
Fig.~\ref{fig:spectralflow}(a) also exposes the mechanism behind the
pinning of the edge occupation established in Sec.~\ref{sec:defects}:
levels of opposite inversion parity cross freely while levels of equal
parity repel, and the two members of the edge doublet exchange sectors
between the initial and the final edge phase---the filled member has
$P = +1$ for $J_z < -\sqrt{3}$ but $P = -1$ for $J_z > \sqrt{3}$---because
within each sector the initial edge level merges into a bulk band at the
gap closing while a different level of the same sector detaches to become
the final edge level.  On the armchair ribbon the sweep traverses the
transient edge window of Fig.~\ref{fig:edgephases}(d): throughout the
window the lowest branch retains a gap exponentially small in $N_2$
[Figs.~\ref{fig:spectralflow}(b) and \ref{fig:spectralflow}(c)], and the
width parity established in Sec.~\ref{sec:geometry} is directly visible:
for odd $N_2$ the crossing at $J_z = 0$ remains exact at every momentum
[Fig.~\ref{fig:spectralflow}(b)], whereas for even $N_2$ it is avoided
[Fig.~\ref{fig:spectralflow}(c)].

At the end of the ramp the evolved state is resolved with respect to the
eigenstates of the final Hamiltonian, using the projectors $P_\pm(k)$ and
$P_0(k)$ of Eq.~\eqref{eq:projectors} evaluated at $t = T$.  For the
present protocol the final Hamiltonian lies deep in the $A_z$ phase, where
the zigzag ribbon supports the edge doublet at every momentum while the
armchair ribbon supports none, so that $P_0(k) = 0$ for the armchair
ribbon and for the tubes.  The defect number in sector $k$ is
\begin{equation}
  \mathcal{N}_+(k) := \Tr\bigl[ P_+(k)\, \rho(k) \bigr] ,
  \label{eq:Nplus}
\end{equation}
with $\rho(k)$ the final single-particle correlation matrix, and
analogously $\mathcal{N}_0(k) := \Tr[ P_0(k)\, \rho(k) ]$ for the
edge-doublet occupation.  The intensive counts are the corresponding
Brillouin-zone averages,
\begin{equation}
  \frac{\mathcal{N}_+}{N_x}
  := \frac{1}{N_x} \sum_{k} \mathcal{N}_+(k)
  = \frac{1}{\pi} \int_0^{\pi} dk\, \mathcal{N}_+(k) ,
  \label{eq:Navg}
\end{equation}
where the sum runs over the full Brillouin zone $k = 2\pi p/N_x$ and the
second form uses $\mathcal{N}_+(-k) = \mathcal{N}_+(k)$; $\mathcal{N}_0/N_x$
is defined in the same way.  With this normalization, an occupation of
exactly one fermion in every sector $0 < k < \pi$ gives
$\mathcal{N}_0/N_x = 1$, and the physical number of excitations on a
ribbon of $N_x$ cells is $\mathcal{N}_+/2$, since the reality constraint
pairs the sectors $k$ and $-k$.  Since the
critical point of each transverse chain is traversed at a rate set by
$\tau$ while the level spacing is set by the width, the natural scaling
variable is $N_2/\sqrt{\tau}$, the ratio of the width to the KZ
correlation length $\hat\xi \propto \sqrt{\tau}$ of a critical line with
dynamical exponent $z = 1$ and correlation-length exponent $\nu = 1$.

For the entanglement analysis we bipartition each geometry along the two
inequivalent lattice directions: a \emph{zigzag cut} runs parallel to
$\bfalpha_1$ and severs one row of $z$-links, while an \emph{armchair cut}
runs parallel to $2\bfalpha_2 - \bfalpha_1$ and severs $x$- and $y$-links.  
We monitor the quench-generated entanglement through the late-time
plateaus of two standard measures, the \emph{entanglement entropy} and the
\emph{logarithmic negativity}, taken relative to their initial
ground-state values and denoted $\Delta\mathcal{S}$ and
$\Delta\mathcal{E}$, respectively.
The entanglement entropy $\mathcal{S}$ is the von Neumann entropy of the half-system reduced density matrix.
As a genuinely quantum counterpart we evaluate the fermionic logarithmic
negativity~\cite{Shapourian17a}
\begin{equation}
  \mathcal{E}
  = \ln \bigl\lVert \rho^{R_A} \bigr\rVert_1 ,
  \label{eq:negativity}
\end{equation}
where $\rho^{R_A}$ denotes the partial time-reversal transform of the
density matrix on subsystem $A$ and
$\lVert X \rVert_1 := \Tr\sqrt{X^\dagger X}$ is the trace norm.  Unlike
the entanglement entropy, which quantifies entanglement only for pure
states of the total system, the logarithmic negativity remains a faithful
entanglement measure for mixed states.  Both $\mathcal{S}$ and
$\mathcal{E}$ are computed from the covariance matrix of the Gaussian
state~\cite{Peschel03a,Peschel2009,Shapourian17a}.  As a reference we also
record both quantities for the diagonal (dephased) mixture of the final
state in the eigenbasis of the final Hamiltonian.  The entropy of this
mixture is not an entanglement measure, but it bounds the pure-state
entanglement entropy from above and thereby isolates the contribution of
the residual coherences~\cite{Alba17a}; its negativity, by contrast, is a
proper entanglement measure of the dephased state and reveals which part
of the quench-generated entanglement survives the loss of coherence
(Sec.~\ref{sec:entanglement}).

All results are obtained by exact propagation of the single-particle
covariance matrix with a fourth-order commutator-free Magnus
integrator~\cite{Blanes06a,Alvermann11a}; related commutator-free
propagators have recently been analyzed in the context of quantum
simulation~\cite{Ikeda23b,Moreno-Casares24a}.
Particle-number conservation
within each momentum sector is exact and serves as a stringent check on the
propagation; system sizes reach $N_2 = 32$ at $N_x = 64$ ($N_x = 121$ for
the tubes); the results are insensitive to the particular choice of $N_x$.
All computations are implemented in a Wolfram Language package built on the Q3 framework~\cite{Choi22a}.

\section{Results: Defect Production}
\label{sec:defects}

\begin{figure}
    \centering
    \includegraphics[width=80mm]{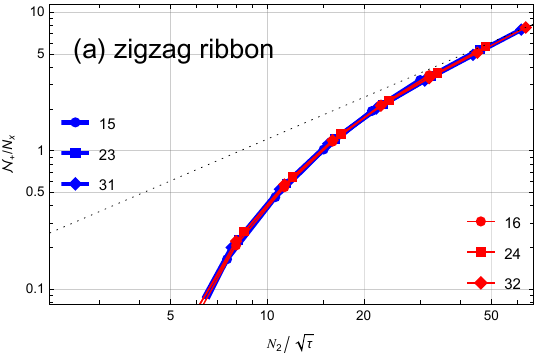}
    \includegraphics[width=80mm]{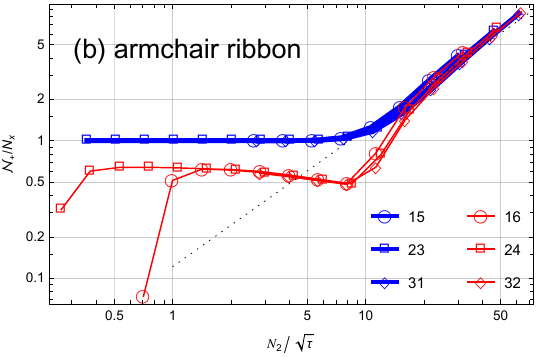}
    \includegraphics[width=80mm]{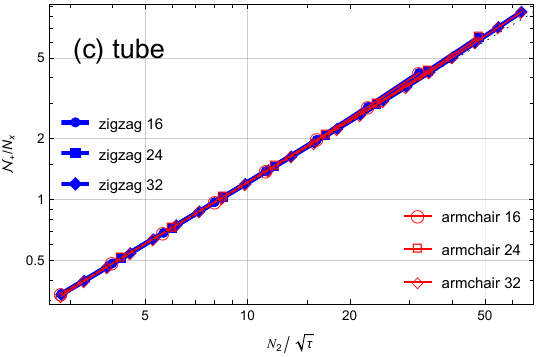}
    \caption{Finite-size scaling of the number of defects generated by the
      quench for $J_x = 1$, $J_y = 2$, and $N_x = 64$ ($N_x = 121$ for
      tubes).  The numbers in the plot legends indicate $N_2$, the number of
      dimer lines in the Kitaev ribbon or tube.  (a)~Zigzag ribbon: the data
      for all widths collapse onto a single function of $N_2/\sqrt{\tau}$,
      following the Kibble--Zurek law (dotted line) for
      $N_2/\sqrt{\tau} \gtrsim 20$ and falling off exponentially below it.  
      (b)~Armchair ribbon: odd widths (open blue symbols,
      thick lines) are pinned at $\mathcal{N}_+/N_x = 1$ by the exact zero-energy
      crossing at $J_z = 0$,
      while even widths (open red symbols, thin lines) settle on the transient plateau
      $\mathcal{N}_+/N_x \approx 0.5$; all widths join the Kibble--Zurek branch at
      $N_2/\sqrt{\tau} \gtrsim 20$.  (c)~Tube: the zigzag-wrapped (filled
      blue symbols) and armchair-wrapped (open red symbols) tubes collapse
      onto one Kibble--Zurek line, indistinguishable from each other at all
      sizes shown.}
    \label{fig:DefectsScaling}
\end{figure}

Figure~\ref{fig:DefectsScaling} summarizes the defect production on the
zigzag [Fig.~\ref{fig:DefectsScaling}(a)] and armchair
[Fig.~\ref{fig:DefectsScaling}(b)] ribbons, our central results, together
with the tubes [Fig.~\ref{fig:DefectsScaling}(c)], which provide the
boundary-free reference against which the ribbon data are compared.
The tubes realize the ideal KZ scenario:
each momentum sector contributes a transverse chain
whose gap closes linearly in time at isolated points of the transverse
momentum grid, and the Brillouin-zone average converts the resulting
Landau--Zener excitation probabilities into the critical-line KZ
law~\cite{Sengupta08a,Mondal08a},
\begin{equation}
  \frac{\mathcal{N}_+}{N_x} = C\, \frac{N_2}{\sqrt{\tau}} ,
  \label{eq:KZlaw}
\end{equation}
with a coupling-dependent constant $C$.  The data for both wrapping
directions and all circumferences collapse onto the single
line~\eqref{eq:KZlaw} over the entire range of $N_2/\sqrt{\tau}$
studied.  The insensitivity to the wrapping direction is not accidental:
the fixed-$k$ mode ladders of the two wrappings coincide, mode for mode,
after the reflection $k \to \pi - k$ whenever $N_2$ is a multiple of four,
so the equality of the two data sets in
Fig.~\ref{fig:DefectsScaling}(c) is an identity rather than an asymptotic
statement.

The zigzag ribbon [Fig.~\ref{fig:DefectsScaling}(a)] follows the same
scaling function for $N_2/\sqrt{\tau} \gtrsim 20$ but departs from it
exponentially at small $N_2/\sqrt{\tau}$.  
The immediate reason for the cutoff is spectral: at fixed $k$ the open
transverse chain has
$\det D_{\mathrm{zz}} = ( J_x + J_y \ee^{-\ii k} )^{N_2} \neq 0$, so its gap
never closes at finite width, and the minimum gap over the entire sweep is
bounded below by the algebraic floor
\begin{equation}
  \Delta_* \simeq |J_x - J_y| \sin\frac{\pi}{N_2+1} ,
  \label{eq:floor}
\end{equation}
attained at $k = \pi$.  A nonzero floor sets a single global adiabatic
threshold $\tau_{\mathrm{ad}} \propto \Delta_*^{-2} \sim N_2^2$ beyond which
every momentum sector freezes simultaneously, which is precisely the
observed cutoff at $\sqrt{\tau} \sim N_2$.  Since the prefactor of the
floor is set by the anisotropy $|J_x - J_y|$, the position of the cutoff,
and with it the onset of the KZ regime, depends on the couplings; the
value $N_2/\sqrt{\tau} \approx 20$ quoted here and below refers to the
ratio $J_y/J_x = 2$ used throughout.  The deeper origin of the
floor, however, is topological and traces back to the zero-energy edge
modes required by the bulk--boundary correspondence.  On the tube the gap
closings are exact: at isolated points of the transverse momentum grid a
mode crosses zero energy and feeds the KZ law~\eqref{eq:KZlaw}.  On the
ribbon the open boundary converts the gap closing into edge-mode
formation: as the transverse chain is driven into its topological
phase~\eqref{eq:zzedge}, the lowest level detaches from the bulk band and
becomes the hybridized doublet of Majorana edge modes, approaching zero
energy only exponentially in the width
[Fig.~\ref{fig:spectralflow}(a)], while the bulk levels left behind
retain the algebraic floor~\eqref{eq:floor}.  The exact critical crossing
of the periodic geometry is thus traded for a zero-energy edge channel,
and a finite zigzag ribbon is \emph{less} defect-prone than a tube of
equal width not in spite of its topological edge modes but because of
them.

The fate of the edge channel itself is fixed by symmetry.  The occupation
of the final edge doublet is insensitive to the ramp rate altogether: we
find
\begin{equation}
  \frac{\mathcal{N}_0}{N_x} = 1
  \label{eq:edgepinning}
\end{equation}
identically, to twelve digits, for every width and every $\tau$.  The
pinning is an effect of inversion parity: within each parity sector of the
gauge-transformed chain~\eqref{eq:parityblocks} the mid-gap level is
nondegenerate and connects
adiabatically to exactly one member of the final edge doublet, so precisely
one fermion is deposited on the edge regardless of how fast the bulk is
excited.  The boundary thus shapes the zigzag quench twice over: the
formation of the zero-energy edge doublet absorbs the critical gap
closing and renders the bulk adiabatic, while inversion parity fixes the
edge-mode occupation deterministically.  Since the occupation of the final
edge doublet is thus pinned, the discussion below focuses on the band
occupation $\mathcal{N}_+/N_x$.

The armchair ribbon [Fig.~\ref{fig:DefectsScaling}(b)] shows the opposite
behavior: neither an exponential cutoff nor a complete collapse.  For
$N_2/\sqrt{\tau} \gtrsim 20$ all widths follow the KZ
branch~\eqref{eq:KZlaw}.  Below it, the defect number saturates at a
width-independent value set by the width parity, and the two plateaus
differ in character: odd widths are pinned at $\mathcal{N}_+/N_x = 1$,
exactly and with no adiabatic threshold at any ramp rate examined, whereas
even widths settle on an approximate and transient plateau
$\mathcal{N}_+/N_x \approx 0.5$ that is decorated, for the smaller widths,
by St\"uckelberg oscillations.  Here, too, zero-energy edge modes are at work, now in
their transient guise inside the edge window of
Figs.~\ref{fig:edgephases}(b) and \ref{fig:edgephases}(d).  Inside the
window the Bloch symbol of the transverse chain is topologically wound,
and the bulk--boundary correspondence accordingly furnishes a pair of
Majorana edge modes at \emph{every} momentum; across the finite width
they hybridize into a single low-energy branch, with a singular value
exponentially small in $N_2$, seeded in the limit $k \to 0$ by the
inversion-parity-protected crossings of Eq.~\eqref{eq:crossings}.  For
any ramp with $\tau \ll \ee^{\mathcal{O}(N_2)}$ this branch is traversed
completely diabatically and acts as an exact level crossing, depositing an
excitation of order unity per momentum sector; for odd $N_2$ the exact
zero-energy crossing at $J_z = 0$ persists at every $k$, which fixes the
exact value $\mathcal{N}_+/N_x = 1$.  The resulting two-channel form
\begin{equation}
  \frac{\mathcal{N}_+}{N_x} \approx C\, \frac{N_2}{\sqrt{\tau}} + n_{\mathrm{edge}} ,
  \label{eq:twochannel}
\end{equation}
with $n_{\mathrm{edge}}$ the width-parity-dependent edge contribution
($n_{\mathrm{edge}} = 1$ for odd and $n_{\mathrm{edge}} \approx 0.5$ for
even widths),
places the crossover between the plateau and the KZ branch at
$N_2/\sqrt{\tau} \approx n_{\mathrm{edge}}/C$, consistent with
Fig.~\ref{fig:DefectsScaling}(b).  Unlike the odd-width value, the
even-width plateau is not protected by an exact crossing: it results from
the diabatic passage through the exponentially small avoided crossings of
the wound branch and is therefore transient, persisting up to a collapse
time that grows exponentially with the width, in proportion to the
inverse of the minimal gap of that branch.  At fixed width the plateau
thus eventually decays, and the slow, non-monotonic approach visible for
the smaller widths in Fig.~\ref{fig:DefectsScaling}(b) is St\"uckelberg
interference between the successive avoided crossings of
Eq.~\eqref{eq:crossings}; at fixed $N_2/\sqrt{\tau}$, by contrast, the
large-width limit is the plateau itself.  In this precise sense the
armchair termination converts a subset of the bulk KZ channels into
edge-mode channels that are immune to the adiabatic limit; the
parity-dependent plateaus of Fig.~\ref{fig:DefectsScaling}(b) are a direct dynamical
fingerprint of the zero-energy edge modes dictated by the bulk--boundary
correspondence, not a finite-size artifact.

\section{Results: Entanglement Generation}
\label{sec:entanglement}

\begin{figure*}
\includegraphics[width=70mm]{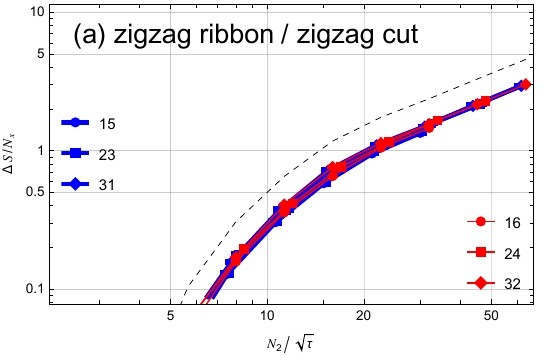}
\includegraphics[width=70mm]{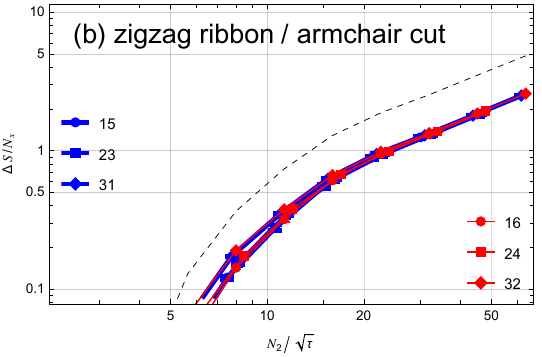} \\
\includegraphics[width=70mm]{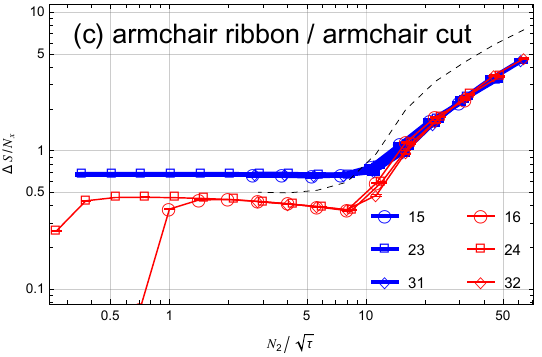}
\includegraphics[width=70mm]{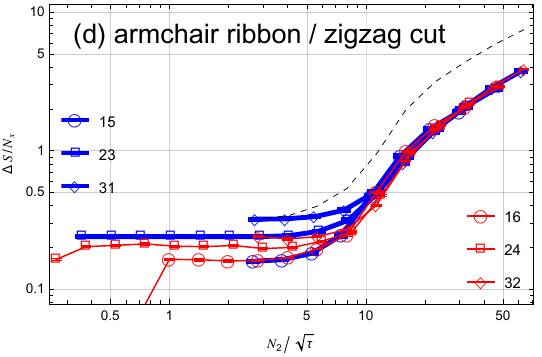} \\
\includegraphics[width=70mm]{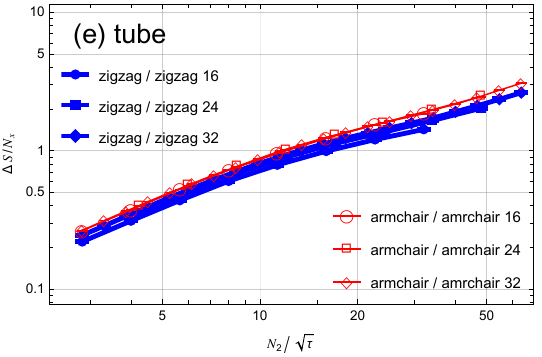}
\includegraphics[width=70mm]{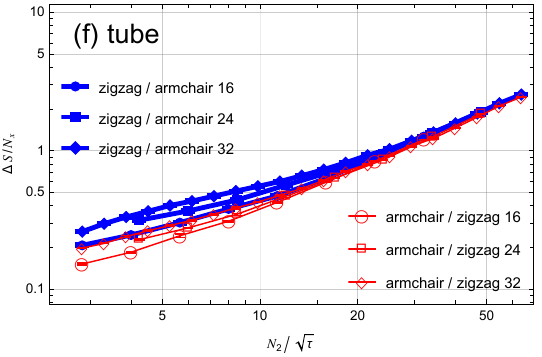}
\caption{Finite-size scaling of the quench-generated entanglement entropy
  of Kitaev ribbons and tubes across longitudinal and transverse cuts.
  Parameters: $J_x = 1$ and $J_y = 2$, $N_x = 64$ ($121$ for tubes).  Panel
  labels give the geometry and the cut orientation; symbol and line
  conventions follow Fig.~\ref{fig:DefectsScaling}.  The dashed lines in
  panels (a)--(d) indicate the entanglement entropy of the diagonal mixture
  in the eigenenergy basis.}
    \label{fig:EntropyScaling}
\end{figure*}

\begin{figure*}
\includegraphics[width=70mm]{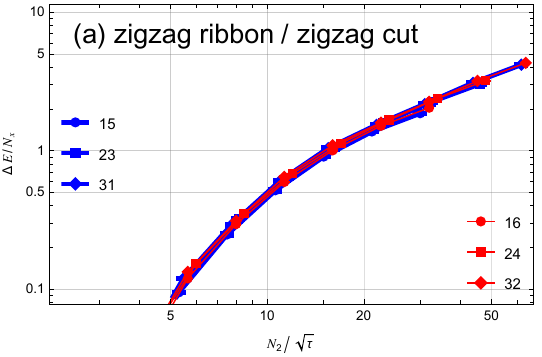}
\includegraphics[width=70mm]{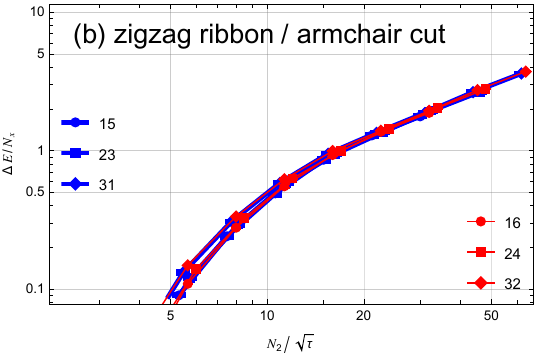} \\
\includegraphics[width=70mm]{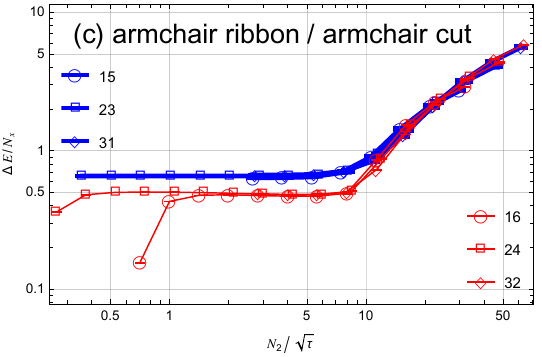}
\includegraphics[width=70mm]{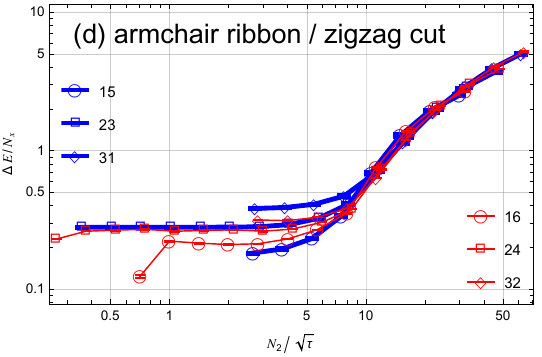} \\
\includegraphics[width=70mm]{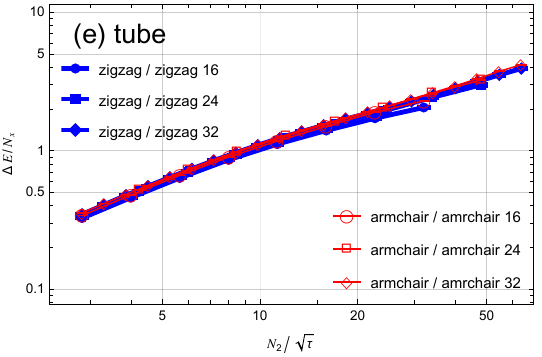}
\includegraphics[width=70mm]{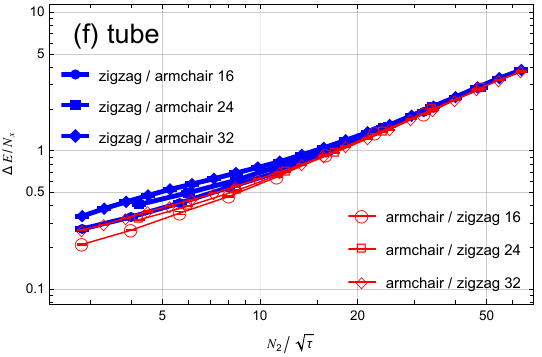}
\caption{Finite-size scaling of the quench-generated logarithmic
  entanglement negativity of Kitaev ribbons and tubes across longitudinal
  and transverse cuts.  Parameters: $J_x = 1$ and $J_y = 2$, $N_x = 64$
  ($121$ for tubes).  Panel labels give the geometry and the cut
  orientation; symbol and line conventions follow
  Fig.~\ref{fig:DefectsScaling}.}
    \label{fig:NegativityScaling}
\end{figure*}

\begin{figure}
\includegraphics[width=70mm]{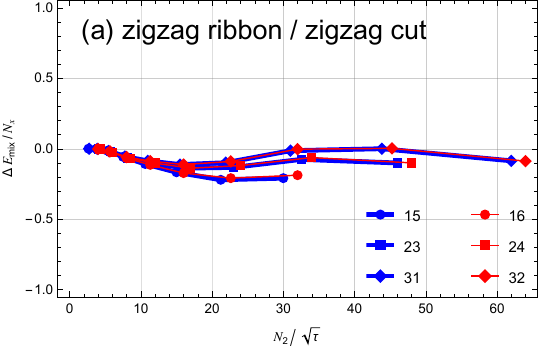} \\
\includegraphics[width=70mm]{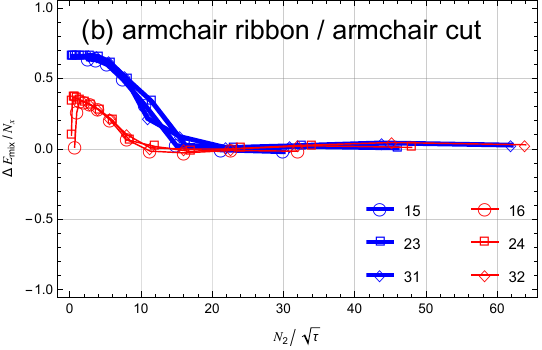} \\
\caption{Quench-generated logarithmic negativity of the diagonal
  (dephased) mixture of the final state, $\Delta\mathcal{E}_{\mathrm{mix}}/N_x$,
  for (a)~the zigzag ribbon across the zigzag cut and (b)~the armchair
  ribbon across the armchair cut, i.e., across the longitudinal cut
  parallel to the edges in both cases; to be compared with the full-state
  negativity in Figs.~\ref{fig:NegativityScaling}(a) and
  \ref{fig:NegativityScaling}(c) and with the diagonal-mixture entropy
  (dashed lines) in Figs.~\ref{fig:EntropyScaling}(a) and
  \ref{fig:EntropyScaling}(c).  Parameters: $J_x = 1$ and $J_y = 2$,
  $N_x = 64$.  Symbol and line conventions follow
  Fig.~\ref{fig:DefectsScaling}.}
    \label{fig:NegativityMixture}
\end{figure}

Figure~\ref{fig:EntropyScaling} presents the quench-generated entanglement
entropy per unit length, $\Delta\mathcal{S}/N_x$, for the two ribbons and the tube,
across both cut orientations, Fig.~\ref{fig:NegativityScaling} the
corresponding fermionic logarithmic negativity $\Delta\mathcal{E}/N_x$, and
Fig.~\ref{fig:NegativityMixture} the negativity
$\Delta\mathcal{E}_{\mathrm{mix}}/N_x$ of the dephased final state.
The entanglement inherits the complete scaling structure of the defect
production, including its edge-mode signatures, as anticipated for integrable systems in which entropy is
carried ballistically by the quasiparticle pairs created during the
ramp~\cite{Calabrese05a,Cincio07a,Alba17a}.

On the tube [Figs.~\ref{fig:EntropyScaling}(e) and
\ref{fig:EntropyScaling}(f)] the entanglement production is extensive and
collapses onto a single function of $N_2/\sqrt{\tau}$ for both wrapping
directions and both cut orientations; cuts parallel and perpendicular to
the wrapping differ only by a small amplitude that diminishes with
increasing $N_2/\sqrt{\tau}$.  On the zigzag ribbon [Figs.~\ref{fig:EntropyScaling}(a)
and \ref{fig:EntropyScaling}(b)] the collapse is equally clean and the
entropy is suppressed below $N_2/\sqrt{\tau} \approx 20$ in parallel with
the defect
cutoff of Fig.~\ref{fig:DefectsScaling}(a); we find no residual entropy
plateau at slow ramps for either cut.  This absence is itself an
edge-mode signature: the edge doublet is occupied by exactly one fermion
both before and after the ramp~\eqref{eq:edgepinning}, so the zero-energy
edge modes, while governing the adiabatic cutoff, deposit no
quench-generated entanglement of their own.  Throughout the scaling
regime the measured plateau entropy lies below the diagonal-mixture
reference (dashed lines) by a factor approaching two, which quantifies the
share of the correlations retained in the coherences between the members of
each quasiparticle pair; a dephased (classical) mixture of the same mode
occupations would carry up to twice the entropy actually generated.

The armchair ribbon [Figs.~\ref{fig:EntropyScaling}(c) and
\ref{fig:EntropyScaling}(d)] reproduces, in the entanglement, the
width-parity-dependent structure of its defect production.  At small
$N_2/\sqrt{\tau}$ the entropy is dominated by the transient zero-energy
edge channel: odd
widths form a ramp-rate-independent plateau, and even widths a lower,
transient one, with the same non-monotonic St\"uckelberg residue at the
smallest widths; both classes join the extensive KZ branch for
$N_2/\sqrt{\tau} \gtrsim 20$.  The plateaus are
more pronounced for the armchair cut, which runs parallel to the edges
hosting the transient edge modes, than for the zigzag cut.  The logarithmic
negativity (Fig.~\ref{fig:NegativityScaling}) tracks the entropy
panel by panel---the same collapse and cutoff on the zigzag ribbon, the
same width-parity-dependent plateaus on the armchair ribbon, and the same
wrapping-independent collapse on the tube---which certifies that the
quench-generated correlations quantified by $\Delta\mathcal{S}$ are genuinely
quantum rather than an artifact of classical mixing~\cite{Shapourian17a}.
In particular, the width-parity-dependent plateaus persist in the
negativity, which identifies the correlations generated by the transient
zero-energy edge modes as genuine quantum entanglement.

The diagonal mixture makes this distinction sharper still.
Figure~\ref{fig:NegativityMixture} shows the quench-generated negativity
of the dephased final state for the two ribbons across the longitudinal
cut, to be compared with the dashed lines of
Figs.~\ref{fig:EntropyScaling}(a) and \ref{fig:EntropyScaling}(c).  On
the zigzag ribbon [Fig.~\ref{fig:NegativityMixture}(a)] the dephased
negativity vanishes over the entire range of $N_2/\sqrt{\tau}$, within
$\pm 0.2$ per unit length and dipping slightly below zero around the
crossover, whereas the entropy of the same mixture exceeds the pure-state
value by up to a factor of two.  The excess entropy of the mixture is
therefore classical mixedness rather than entanglement, and the
entanglement generated in the KZ regime resides entirely in the coherences
between the members of each quasiparticle pair, whose Landau--Zener
occupations are fractional.  The armchair ribbon
[Fig.~\ref{fig:NegativityMixture}(b)] separates the two regimes: for
$N_2/\sqrt{\tau} \gtrsim 20$ the dephased negativity vanishes as on the
zigzag ribbon, but in the plateau regime most of it survives
dephasing---about $0.65$ per unit length for odd widths against the
pure-state plateau of about $0.7$, and a correspondingly reduced transient
value for even widths.  The edge channel is traversed completely
diabatically, so the final occupations it leaves behind are close to
integers and the dephased state remains nearly pure within that channel;
the plateau entanglement is thus borne by the populations of final
eigenmodes straddling the cut rather than by pair coherences, which makes
it, unlike the KZ-regime entanglement, robust against dephasing in the
energy eigenbasis.

\section{Summary}
\label{sec:summary}

We have analyzed defect production and entanglement generation in the
flux-free sector of the Kitaev honeycomb model driven across its
topological transition by a linear ramp of $J_z$, comparing tubes with
zigzag and armchair ribbons at fixed anisotropy $J_y/J_x = 2$.  On the
tube the defect density follows the critical-line Kibble--Zurek law,
Eq.~\eqref{eq:KZlaw}, collapsing onto the single variable
$N_2/\sqrt{\tau}$ independently of the wrapping direction.  The
termination reshapes this picture qualitatively, and it does so in
opposite directions for the two edge orientations.  The zigzag ribbon,
whose transverse chains retain the algebraic gap floor of
Eq.~\eqref{eq:floor}, becomes globally adiabatic at
$\sqrt{\tau} \gtrsim N_2$ and produces exponentially fewer defects than
the corresponding tube, while inversion parity pins the edge-doublet
occupation to exactly one fermion.  The armchair ribbon supports a
topologically wound transient edge channel, seeded by the
inversion-parity-protected crossings of Eq.~\eqref{eq:crossings}, that
acts as an exact level crossing for any subexponential ramp and leaves
behind a defect plateau set by the even--odd parity of the width.  The entanglement
entropy and the fermionic logarithmic negativity, for both inequivalent
cuts, inherit this entire structure and confirm its quantum character.

Beyond the specific model, the results identify the spectral criterion
that decides the fate of a driven channel: families of transverse modes
whose gaps possess zeros as a function of the conserved momentum
contribute the universal KZ power law, whereas families with a gap floor
freeze out at a single global threshold---algebraic floors (zigzag) make
the threshold reachable, exponential floors (armchair) push it beyond any
practical ramp time and leave the channel at a parity-dependent plateau.  Since the edge
orientation of a sample selects between these behaviors, slow-ramp defect
counting acts as a sensitive probe of the boundary topology, a scenario
directly relevant to current realizations of the Kitaev model on
programmable quantum processors~\cite{Sun23a,Evered2025,Will2025}.  Extending the
analysis beyond the flux-free free-fermion sector, to include gauge-flux
excitations and their interplay with the boundary channels, remains an
open problem.

\begin{acknowledgments}
H.C. and M.-S.C.\ were supported by the National Research Foundation of Korea (Grant Nos.~RS-2023-NR068116, RS-2023-NR057243, 2023R1A2C1005588, RS-2024-00432563, RS-2024-00404854).
A.M.\ was supported by the IRA Programme, project no. FENG.02.01-IP.05-0006/23, financed by
the FENG program 2021--2027, Priority FENG.02, Measure FENG.02.01., with the support of
the FNP.
We acknowledge the use of a computational server financed by the Foundation for Polish Science (IRAP project, ICTQT, contract no. 2018/MAB/5/AS-1, co-financed by EU within Smart Growth Operational Programme).
\end{acknowledgments}

\bibliography{References}

\end{document}